\documentclass[aps,pra,twocolumn,superscriptaddress,showpacs]{revtex4}
\usepackage{amsmath}
\usepackage{graphicx}
\usepackage{subfigure}
\usepackage{bm}

\begin{document}

\title{A possibility to measure elastic photon--photon scattering in vacuum}

\author{Daniel Eriksson} 
\affiliation{Department of Physics, Ume{\aa} University, SE--901
  87 Ume{\aa}, Sweden}

\author{Gert Brodin} 
\affiliation{Department of Physics, Ume{\aa} University, SE--901
  87 Ume{\aa}, Sweden}

\author{Mattias Marklund}
\altaffiliation[Also at: ]{Institut f\"ur Theoretische Physik IV,
  Ruhr-Universit\"at Bochum, D--44780 Bochum, Germany}
\affiliation{Department of Electromagnetics, Chalmers University of
  Technology, SE--412 61 G\"oteborg, Sweden}

\author{Lennart Stenflo}
\affiliation{Department of Physics, Ume{\aa} University, SE--901
  87 Ume{\aa}, Sweden}

\date{\today}

\begin{abstract}
Photon--photon scattering in vacuum due to the interaction with virtual
electron-positron pairs is a consequence of quantum electrodynamics. A way
for detecting this phenomenon has been devised based on interacting modes
generated in microwave waveguides or cavities [G.\ Brodin, M.\ Marklund and
L.\ Stenflo, Phys.\ Rev.\ Lett.\ \textbf{87} 171801 (2001)]. Here we
materialize these ideas, suggest a concrete cavity geometry, make
quantitative estimates and propose experimental details. It is found that
detection of photon-photon scattering can be within the reach of present day
technology.
\end{abstract}


\maketitle

\section{Introduction}

Classically, photon--photon scattering does not take place in vacuum.
However, according to quantum electrodynamics (QED), such a process may
occur owing to the interaction with virtual electron--positron pairs. An
effective field theory containing only the electromagnetic fields can be
formulated in terms of the Heisenberg--Euler Lagrangian \cite
{Heisenberg-Euler,Schwinger}, which is valid for field strengths below the
QED critical field $10^{18}\,\mathrm{V\,m^{-1}}$ and for wavelengths shorter
than the Compton wavelength $10^{-12}\,\mathrm{m}$ \cite
{general,general2,general3}. Several suggestions to detect photon-photon
scattering in laboratories have been made \cite{Older,Older2,Older3}, and
the recent increase in available laser intensities has stimulated various
schemes \cite{Recent,Recent2,Recent3,Recent4,Recent5,Recent6}. In Ref. \cite
{Brodin-Marklund-Stenflo} we suggested an alternative method, using the
significantly weaker (but still strong) fields that can be confined in a
microwave cavity. The main advantage of adopting a cavity is that we can
achieve a \textit{resonant} interaction between the eigenmodes in a large
volume, leading to an excited signal at a new eigenfrequency.

In the present paper we materialize the proposal made in Ref. \cite
{Brodin-Marklund-Stenflo} in several ways. Various concrete geometries
fulfilling all resonance and frequency matching conditions in a cavity are
devised, and the coupling between the pump modes and the excited mode is
evaluated for a rectangular prism and a cylindrical geometry. The amplitude
of the excited mode is then determined in terms of the quality factor of the
cavity and the field strengths of the pump modes. Comparing with the
performance reached in existing superconducting niobium cavities, it turns
out that the cavity key parameters, namely the quality factor and the
allowed field strength (before field emission and/or superconductivity break
down) can reach values which allow for several photons in the excited mode.
To be able to detect the very weak excited signal in the presence of the
pump waves, we propose a ''cavity filtering geometry''. Finite element
calculations are made in order to demonstrate a geometry that fulfills the
relevant resonance and frequency matching conditions. The estimated signal
in the filtered region corresponds to roughly 30 microwave photons/$\mathrm{m%
}^{3}$. The results of Refs.\ \cite{Q-factor1} and \cite{NOGUES99}, where
effects of single microwave photons confined in cavities are measured using
sensitive techniques involving transitions of Rydberg atoms, suggest that
detection of the estimated signal is possible. Hence we suggest that elastic
photon-photon scattering can be observed with present day technology.

\section{Basic equations and principles of calculation}

Photon-photon scattering due to the interaction with virtual electron
positron pairs can be described by the Heisenberg--Euler Lagrangian \cite
{Heisenberg-Euler} 
\begin{equation}
L=\varepsilon _{0}F+\kappa \varepsilon _{0}^{2}\left[ 4F^{2}+7G^{2}\right]
\label{eq:lagrangian}
\end{equation}
where $F=(E^{2}-c^{2}B^{2})/2$ and $G=c\mathbf{E}\cdot \mathbf{B}$. Here $%
\kappa \equiv 2\alpha ^{2}\hbar ^{3}/45m_{e}^{4}c^{5}\approx 1.63\times
10^{-30}\,\mathrm{m}\mathrm{s}^{2}/\mathrm{kg}$, $\alpha $ is the
fine-structure constant, $\hbar $ Planck's constant, $m_{e}$ the electron
mass, and $c$ the velocity of light in vacuum. The last terms in (\ref
{eq:lagrangian}) represent the effects of vacuum polarization and
magnetization. The QED corrected Maxwell's vacuum equations can then be
written in their classical form using $\mathbf{D}=\varepsilon _{0}\mathbf{E}+%
\mathbf{P}\ $and $\mathbf{H}=\mathbf{B/}\mu _{0}-\mathbf{M}$ \ where $%
\mathbf{P}$ and $\mathbf{M}$ are of third order in the field amplitudes $%
\mathbf{E}$ and $\mathbf{B}$, and $\mu _{0}=1/c^{2}\varepsilon _{0}$.
Expressions for $\mathbf{P}$ and $\mathbf{M}$ can be found in, for example,
Refs.\ \cite
{general,general2,general3,Older,Older2,Older3,Recent,Recent2,Recent3,Recent4,Recent5,Brodin-Marklund-Stenflo}%
.

There are several reasons for considering wave interactions in cavities:

\begin{enumerate}
\item  We can benefit from coherent resonant interactions. In contrast, the nonlinear
coupling vanishes for parallell plane waves in an unbounded medium. However,
we note that the presence of an inhomogeneous background magnetic field can
also be responsible for a nonzero effect, see Ref. \cite{Recent6} .

\item  The growth of the new mode will not be saturated by convection out of the
interaction region.

\item  The techniques for detecting small signals in such cavities are
very well developed, see e.g. Refs \cite{Q-factor1,NOGUES99}.

\end{enumerate}

Calculations of the coupling strength between various eigenmodes can be made
including the nonlinear \ polarization and magnetization, see e.g. Ref. \cite
{Brodin-Marklund-Stenflo}. However, a more convenient approach, which gives
the same result, starts directly with the Lagrangian density (\ref
{eq:lagrangian}).

Our general procedure for finding the cavity eigenmode coupling and the
saturated amplitude of the excited mode, to be applied in sections III and
IV, can be summarized as follows: 

1.~We determine the \textit{linear eigenmodes} of the cavity using the
standard Maxwell vacuum equations together with the boundary conditions for
infinitely conducting walls, and express all field components in terms of
the vector potential amplitude. 

2.~Next we choose resonant eigenmodes fulfilling frequency matching
conditions. For two interacting initial pump modes (indices 1 and 2) with
distinct frequencies $\omega _{1}$ and $\omega _{2}$, the possible frequency
matching choices, corresponding to a cubic nonlinearity, are $\pm (2\omega
_{1}\pm \omega _{2})$ and $\pm (\omega _{1}\pm 2\omega _{2})$. For
definiteness we concentrate on the choice 
\begin{equation}
\omega _{3}=2\omega _{1}-\omega _{2}  \label{frequency matching}
\end{equation}
in all examples. Since the eigenfrequencies are determined by the geometry,
we note that the condition (\ref{frequency matching}) gives a design
requirement involving the dimensions of the cavity. In this context it can
be noted that there are several reasons for using two pump waves rather than
one, that in principle could excite a mode with three times the frequency of
the pump mode. Firstly, for two pump modes there are generally better
possibilities to vary the parameters in the set-up in order to optimize the
output of the excited mode. Secondly, for a cylindrical geometry and a
single pump mode, the design requirements following from the frequency
matching results in eigenmodes that have a too weak nonlinear coupling.
Finally, there is a general tendency to get stronger nonlinear coupling when
the pump modes and the excited mode have rather close frequencies.

3.~We then perform a variation of the amplitude of the eigenmodes and let $%
\delta \!\int L\,d^{3}rdt=0$ in order to obtain the mode-coupling equations
directly from the Lagrangian density (\ref{eq:lagrangian}). The evolution
equation for each eigenmode is obtained by expressing the Lagrangian in
terms of the potential and varying the corresponding vector potential
amplitude. The lowest order linear terms then vanish, since the dispersion
relation of each mode is fulfilled. For the terms that are quadratic in the
fields we must thus take into account that the amplitude has a weak time
dependence when making the amplitude variation. However, for the QED
correction terms in the Lagrangian the time dependence of the amplitudes can
be neglected. 

4.~In the absence of dissipation, the equations now obtained imply steady
growth of mode 3, until the energy of that mode is comparable to that of the
pump modes. However, when some damping mechanism is present (e.g. due to a
finite conductivity of the cavity walls) the amplitude saturates at a level
where the mode-coupling growth balances the dissipation of the excited mode.
This effect can be easily included by adding a phenomenological damping term
in the evolution equation, whose value is estimated by comparing with
quality factors currently reached in superconducting niobium cavities. 

\section{Rectangular prism geometry}

We start by considering a rectangular prism cavity, with one of its corners
in the origin, and the opposite corner with coordinates $(x_{0},y_{0},z_{0})$%
. In practice we are interested in a shape where $z_{0}\gg x_{0},y_{0}$ but
this assumption will not be used in the calculations. We let the large
amplitude pump modes have vector potentials of the form 
\begin{equation}
\mathbf{A}_{1}=A_{1}\sin \left( \frac{\pi x}{x_{0}}\right) \sin \left( \frac{%
n_{1}\pi z}{z_{0}}\right) \exp (-\mathrm{i}\omega _{1}t)\widehat{\mathbf{y}}+%
\mathrm{c.c}  \label{vector-rect1}
\end{equation}
and 
\begin{equation}
\mathbf{A}_{2}=A_{2}\sin \left( \frac{\pi y}{y_{0}}\right) \sin \left( \frac{%
n_{2}\pi z}{z_{0}}\right) \exp (-\mathrm{i}\omega _{2}t)\widehat{\mathbf{x}}+%
\mathrm{c.c.}  \label{vector-rect2}
\end{equation}
where $\mathrm{c.c.\,}$\ denotes complex conjugate, $n_{1,2}=1,2,3...$, and
where we have chosen the radiation gauge such that the scalar potential is
zero. It is easily checked that the corresponding fields (omitting the c.c.) 
\begin{subequations}
\begin{eqnarray}
B_{1z}\!\! &=&\!\!\frac{\pi }{x_{0}}A_{1}\,\cos \left( \frac{\pi x}{x_{0}}%
\right) \sin \left( \frac{n_{1}\pi z}{z_{0}}\right) \exp (-\mathrm{i}\omega
_{1}t),  \label{FPR11} \\
B_{1x}\!\! &=&\!\!-\frac{n_{1}\pi }{z_{0}}A_{1}\sin \left( \frac{\pi x}{x_{0}%
}\right) \cos \left( \frac{n_{1}\pi z}{z_{0}}\right) \exp (-\mathrm{i}\omega
_{1}t),  \label{FPR12} \\
E_{1y}\!\! &=&\!\!\mathrm{i}\omega _{1}A_{1}\sin \left( \frac{\pi x}{x_{0}}%
\right) \sin \left( \frac{n_{1}\pi z}{z_{0}}\right) \exp (-\mathrm{i}\omega
_{1}t),  \label{FPR13}
\end{eqnarray}
together with $\omega _{1}^{2}=n_{1}^{2}\pi ^{2}c^{2}/z_{0}^{2}+\pi
^{2}c^{2}/x_{0}^{2}$, and 
\end{subequations}
\begin{subequations}
\begin{eqnarray}
B_{2z}\!\! &=&\!\!-\frac{\pi }{y_{0}}A_{2}\,\cos \left( \frac{\pi y}{y_{0}}%
\right) \sin \left( \frac{n_{2}\pi z}{z_{0}}\right) \exp (-\mathrm{i}\omega
_{2}t),  \label{FPR21} \\
B_{2y}\!\! &=&\!\!\frac{n_{2}\pi }{z_{0}}A_{2}\sin \left( \frac{\pi y}{y_{0}}%
\right) \cos \left( \frac{n_{2}\pi z}{z_{0}}\right) \exp (-\mathrm{i}\omega
_{2}t),  \label{FPR22} \\
E_{2x}\!\! &=&\!\!\mathrm{i}\omega _{2}A_{2}\sin \left( \frac{\pi y}{y_{0}}%
\right) \sin \left( \frac{n_{2}\pi z}{z_{0}}\right) \exp (-\mathrm{i}\omega
_{2}t),  \label{FPR23}
\end{eqnarray}
together with $\omega _{2}^{2}=n_{2}^{2}\pi ^{2}c^{2}/z_{0}^{2}+\pi
^{2}c^{2}/y_{0}^{2}$ \ are proper eigenmodes fulfilling Maxwells equations
and the standard boundary conditions. Similarly we assume that the mode to
be excited can be described by a vector potential 
\end{subequations}
\begin{equation}
\mathbf{A}_{3}=A_{3}\sin \left( \frac{\pi y}{y_{0}}\right) \sin \left( \frac{%
n_{3}\pi z}{z_{0}}\right) \exp (-\mathrm{i}\omega _{3}t)\widehat{\mathbf{x}}+%
\mathrm{c.c.}  \label{vector-rect3}
\end{equation}
where $\omega _{3}^{2}=n_{3}^{2}\pi ^{2}c^{2}/z_{0}^{2}+\pi
^{2}c^{2}/y_{0}^{2}$, in which case we get fields of the same form as in
Eqs. (\ref{FPR21})--(\ref{FPR23}).

Next we turn to point three in the scheme of the previous section. As noted
above, when performing the variations $\delta A_{3}^{\ast }$, the lowest
order terms proportional to $\delta A_{3}^{\ast }A_{3}$ vanish due to the
dispersion relation, and we need to include terms due to the time dependence
of the amplitude of the type $A_{3}\partial (\delta A_{3}^{\ast })/\partial
t $. For the fourth order QED corrections proportional to $\delta
A_{3}^{\ast } $, only terms proportional to $A_{1}^{2}A_{2}^{\ast }\delta
A_{3}^{\ast }$ survive the time integration, due to the frequency matching (%
\ref{frequency matching}). After some algebra the corresponding evolution
equation for mode 3 reduces to 
\begin{equation}
\frac{dA_{3}}{dt}=-\frac{\mathrm{i}\varepsilon _{0}\kappa \omega _{3}^{3}}{8}%
K_{\mathrm{rec}}A_{1}^{2}A_{2}^{\ast }  \label{Evolution1}
\end{equation}
where the dimensionless coupling coefficient $K_{\mathrm{rec}}$ is 
\begin{widetext}
\begin{eqnarray}
K_{\mathrm{rec}} &=&\frac{\pi ^{2}c^{2}}{\omega _{3}^{4}}\Bigg\{(-,-,+)\bigg[%
\frac{8\pi ^{2}c^{2}}{x_{0}^{2}y_{0}^{2}}+\left( \frac{4}{x_{0}^{2}}+\frac{%
7n_{1}^{2}}{z_{0}^{2}}\right) \omega _{2}\omega _{3}\bigg]+\frac{%
n_{2}n_{3}\pi ^{2}c^{2}}{z_{0}^{2}}\left( \frac{7n_{1}^{2}}{z_{0}^{2}}-\frac{%
3}{x_{0}^{2}}\right)   \notag \\
&&+\frac{7\omega _{1}n_{1}}{z_{0}^{2}}\left( (-,+,-)\omega
_{2}n_{3}(+,-,-)\omega _{3}n_{2}\right) \Bigg\}  \label{Coupling1}
\end{eqnarray}
\end{widetext}
The three different sign alternatives in (\ref{Coupling1}) correspond to the
mode number matching options 
\begin{subequations}
\begin{eqnarray}
2n_{1}-n_{2}+n_{3} &=&0  \label{mode1} \\
2n_{1}+n_{2}-n_{3} &=&0  \label{mode2} \\
2n_{1}-n_{2}-n_{3} &=&0  \label{mode3}
\end{eqnarray}
respectively, which must be fulfilled in order for the coupling to be
nonzero. It is now possible to evaluate the coupling coefficient for
specific mode numbers and geometries consistent with the frequency matching
conditions. As described in the previous section, we could then determine
the saturated amplitude from a balance between mode-coupling growth and
dissipation, where the latter effect can be introduced by simply adding a
phenomenological damping term. However, it turns out that a cylindrical
geometry gives a slightly better performance, and thus we will instead work
out that case in more detail.

\section{Cylindrical geometry}

We consider the case where all eigenmodes are \textrm{TE}-modes with no
angular dependence, with fields that can be derived from the vector
potential 
\end{subequations}
\begin{equation}
\mathbf{A}=AJ_{1}(\rho \beta /a)\sin \left( \frac{n\pi z}{z_{0}}\right) \exp
(-\mathrm{i}\omega t)\widehat{\bm{\varphi }}+\mathrm{c.c.}
\label{Vector-cyl}
\end{equation}
where $a$ is the cylinder radius, $z_{0}$ the length of the cavity, $J_{1}$
the first order Bessel function and $\beta $ one of its zeros. The cylinder
occupies the region $0\leq z\leq z_{0}$ centered around the $z$-axis. We
have here introduced cylindrical coordinates $\rho $ and $z$ as well as the
unit vector $\widehat{\bm{\varphi }}$ in the azimuthal direction. The
corresponding fields are 
\begin{eqnarray}
&&\mathbf{E}=\mathbf{\mathrm{i}\omega }AJ_{1}(\rho \beta /a)\sin \left( 
\frac{n\pi z}{z_{0}}\right) \exp (-\mathrm{i}\omega t)\widehat{\bm{\varphi }}%
+\mathrm{c.c.}  \label{cylfield1} \\
&&\mathbf{B}=\frac{\beta A}{a}J_{0}(\rho \beta /a)\sin \left( \frac{n\pi z}{%
z_{0}}\right) \exp (-\mathrm{i}\omega t)\widehat{\mathbf{z}}-  \notag \\
&&\quad \frac{n\pi }{z_{0}}AJ_{1}(\rho \beta /a)\cos \left( \frac{n\pi z}{%
z_{0}}\right) \exp (-\mathrm{i}\omega t)\widehat{\bm{\rho }}+\mathrm{c.c.}
\label{Cylfield2}
\end{eqnarray}
where the eigenfrequency is $\omega ^{2}=c^{2}[(\beta /a)^{2}+(n\pi
/z_{0})^{2}]$ for all modes $\omega =\omega _{1,2,3}$. We note from the
frequency matching condition $\omega _{3}=2\omega _{1}-\omega _{2}$ that all
of the eigenmodes cannot have the same order of \ their respective $\beta $,
for resonant interaction to occur, and thus we introduce $\beta =\beta
_{1,2,3}$ for the different modes. Proceeding in the same manner as
described in the two previous sections, we obtain after lengthy but
straightforward algebra 
\begin{equation}
\frac{dA_{3}}{dt}=-\frac{\mathrm{i}\varepsilon _{0}\kappa \omega _{3}^{3}}{8}%
K_{\mathrm{cyl}}A_{1}^{2}A_{2}^{\ast }  \label{Evolution2}
\end{equation}
where the cylindrical coupling coefficient $K_{\mathrm{cyl}}$ is 
\begin{widetext}
\begin{eqnarray}
K_{\mathrm{cyl}} &=&\frac{8c^{2}}{\omega _{3}^{4}J_{0}^{2}(\beta _{3})}%
\Bigg\{c^{2}\bigg[\frac{\beta _{1}^{2}\beta _{2}\beta _{3}}{a^{4}}I_{b}-%
\frac{2\pi ^{4}n_{1}^{2}n_{2}n_{3}}{z_{0}^{4}}I_{a}+\frac{\pi ^{2}\beta _{1}%
}{a^{2}z_{0}^{2}}\left( \beta _{1}n_{2}n_{3}(I_{a}+I_{c})+2\beta
_{2}n_{1}n_{3}I_{d}-2\beta _{3}n_{1}n_{2}I_{e}\right) \bigg]  \notag \\
&&\!\!\!\!\!\!-\frac{\beta _{1}}{a^{2}}\left( 2\beta _{3}\omega _{1}\omega
_{2}I_{e}+2\beta _{2}\omega _{1}\omega _{3}I_{d}-\beta _{1}\omega _{2}\omega
_{3}(I_{c}+3I_{a})\right) -\frac{2\pi ^{2}n_{1}I_{a}}{z_{0}^{2}}\left(
n_{3}\omega _{1}\omega _{2}-n_{2}\omega _{1}\omega _{3}-n_{1}\omega
_{2}\omega _{3}\right) \Bigg\}  \label{coupling2}
\end{eqnarray}
\end{widetext}
and the integrals are defined as 
\begin{subequations}
\begin{equation}
I_{a}=\int_{0}^{1}\!\!J_{1}^{2}\left( \beta _{1}u\right) J_{1}\left( \beta
_{2}u\right) J_{1}\left( \beta _{3}u\right) u\,du,  \label{I1}
\end{equation}
\begin{eqnarray}
I_{b} &=&\int_{0}^{1}\!\!\left[ 3J_{0}^{2}\left( \beta _{1}u\right)
+J_{1}^{2}\left( \beta _{1}u\right) \right]  \notag \\
&&\qquad \times J_{0}\left( \beta _{2}u\right) J_{0}\left( \beta
_{3}u\right) u\,du,  \label{I2}
\end{eqnarray}
\begin{equation}
I_{c}=\int_{0}^{1}\!\!J_{0}^{2}\left( \beta _{1}u\right) J_{1}\left( \beta
_{2}u\right) J_{1}\left( \beta _{3}u\right) u\,du,  \label{I3}
\end{equation}
\begin{equation}
I_{d}=\int_{0}^{1}\!\!J_{0}\left( \beta _{1}u\right) J_{1}\left( \beta
_{1}u\right) J_{0}\left( \beta _{2}u\right) J_{1}\left( \beta _{3}u\right)
u\,du,  \label{I4}
\end{equation}
and 
\begin{equation}
I_{e}=\int_{0}^{1}\!\!J_{0}\left( \beta _{1}u\right) J_{1}\left( \beta
_{1}u\right) J_{1}\left( \beta _{2}u\right) J_{0}\left( \beta _{3}u\right)
u\,du.  \label{I5}
\end{equation}
When calculating (\ref{coupling2}) we have assumed the mode number matching $%
n_{3}=2n_{1}+n_{2}$. Equation (\ref{Evolution2}) implies a linear growth of
mode 3, until the backreaction of the pump modes becomes significant. In
reality dissipative mechanisms (e.g. a finite conductivity of the cavity
walls) will put a limit on the excited amplitude. This can be described in a
phenomenological way by substituting $d/dt\rightarrow d/dt-(\omega _{3}/2\pi
Q)$, where $Q$ is the cavity quality factor. The steady state amplitude is
thus 
\end{subequations}
\begin{equation}
A_{3}=\frac{\mathrm{i}\pi QK_{\mathrm{cyl}}}{4}\frac{\omega _{3}^{2}A_{1}^{2}%
}{E_{\mathrm{char}}^{2}}A_{2}^{\ast }  \label{Saturated-amp}
\end{equation}
where we have introduced the characteristic electric field $E_{\mathrm{char}%
}=(\varepsilon _{0}\kappa )^{-1/2}\approx 2.6\times 10^{20}\mathrm{V/m}$ 
\cite{Note1}. Looking at the number of excited photons in the cavity mode $%
N\approx \varepsilon _{0}\int E_{3}E_{3}^{\ast }d^{3}r/\hbar \omega _{3}$ we
finally obtain 
\begin{equation}
N_{\mathrm{QED}}=\frac{\varepsilon _{0}a^{2}z_{0}\pi ^{3}Q^{2}\omega
_{3}^{5}K_{\mathrm{cyl}}^{2}J_{0}^{2}(\beta _{3})\left| A_{1}\right|
^{4}\left| A_{2}\right| ^{2}}{16\hbar E_{\mathrm{char}}^{4}}
\label{photon-numb}
\end{equation}
Before evaluating (\ref{photon-numb}) we need to specify the mode numbers
and the geometry. As an example we let $(n_{1},n_{2},n_{3})=(3,15,21)$
(fulfilling $n_{3}=2n_{1}+n_{2}$)and take $\beta _{2}=$ $\beta _{3}=3.83$,
corresponding to the first zero of $J_{1}$, and $\beta _{1}=7.01$
corresponding to the second zero. This gives us $z_{0}/a=9.53$ through the
frequency matching condition (\ref{frequency matching}) and determines the
frequency relations to $\omega _{3}/\omega _{2}=1.26$ and $\omega
_{3}/\omega _{1}=1.12$. Substituting these values and numerically evaluating
the integrals (\ref{I1})--(\ref{I5}) then gives $K_{\mathrm{cyl}}=3.39$. The
key parameters are the quality factor and the pump field strength. An
advantage with our choice of eigenmodes is that the pump electric field is
zero at the cavity surface, which means that we do not have to worry about
field emission \cite{Field emission}. Instead the pump amplitude is limited
by the surface magnetic field, which needs to be below the critical value
for which the walls cease to be superconducting. From the experimental
results presented in Ref. \cite{Critical field}, we find that the critical
magnetic field for a high pure niobium material can reach $B\approx 0.28$ 
\textrm{Tesla}, at a temperature of the order of or below $1$ $\mathrm{K}$.
If we specify $z_{0}=2.50$ $\mathrm{m}$, we get $\omega _{1}=8.10\times
10^{9}$ $\mathrm{rad/s}$ and $\omega _{2}=7.15\times 10^{9}$ $\mathrm{rad/s}$%
, implying that field levels close to the magnetic surface field condition
correspond to $A_{1}=0.017$ $\mathrm{Vs/m}$, and $A_{2}\approx 0.024$ $%
\mathrm{Vs/m}$. The conductivity of niobium allow for quality factors $%
Q>10^{11}$ for temperatures in the range of interest below $1$ $\mathrm{K}$.
But we note that these high levels have been hard to reach in practice,
although $Q\approx 4\times 10^{10}$ in Ref. \cite{Q-factor1}. We also note
that there has been a tendency to get lower quality factors when applying
stronger fields. However, in Ref. \cite{Q-factor2}, it has been shown that
it is possible to reach surface fields of the order of the critical level
without significant decrease of the Q-factor. Thus adopting $Q=4\times
10^{10}$ and the rest of the parameter values as specified above, we obtain 
\begin{equation}
N_{\mathrm{QED}}\approx 18  \label{Photon-numb2}
\end{equation}
Keeping the cavity at a temperature $T\approx 0.5$ $\mathrm{K}$, thus means
that the number of generated photons in our example is well above the
thermal fluctuation level $N_{\mathrm{th}}=1/[\exp (\hbar \omega
_{3}/kT)-1]\approx 7$, where $k$ is the Boltzmann constant. To get an even
lower thermal fluctuation level, it would be of interest to generate photons
with higher frequencies. However, it is not wise to just scale down the
dimensions used above to get $\hbar \omega _{3}/kT>1$ and $N_{\mathrm{th}%
}\ll 1$, since $N_{\mathrm{QED}}$ decreases too quickly with the cavity
volume. Instead we could consider higher mode numbers $n_{1,2,3}$ and higher
orders of $\beta $ to get a larger excited eigenfrequency fulfilling $\hbar
\omega _{3}/kT>1$ which gives a higher ratio $N_{\mathrm{QED}}/N_{\mathrm{th}%
}$.

On the other hand, detecting a very weak signal in the presence of strong
pump fields might be difficult even for signals well above the thermal
fluctuation level, and thus we will below investigate the possibility of a
cavity geometry that directly filters away the pump signals.

\section{Cavity filtering geometry}

We search for a cavity geometry with the following properties:

\begin{enumerate}
\item  The cavity should be rotationally symmetric

\item  There should be three eigenmodes with frequencies fulfilling (\ref
{frequency matching})

\item  There should be a cylindrical interaction region where the mode
structure resembles that considered in the previous section.

\item  There should be a filtering region with a cross section small enough
to give an exponential decay of the pump modes, but large enough for the
excited eigenmode to propagate.

\item  There should be an entrance region for the pump modes.
\end{enumerate}

We thus make a geometrical design of the type outlined above, see Figs.
1a,b,c. The eigenmodes are then calculated using the method of finite
elements. In general there will be eigenmodes with the desired properties,
except that they will not fulfil Eq.\ (\ref{frequency matching}) for the
specified geometry. However, by repeated calculations varying the length $l$
of the cavity, the mismatch frequency $\delta \omega (l)=\omega
_{3}(l)-2\omega _{1}(l)+\omega _{2}(l)$ gradually approaches zero. The
dimensions of the cavity and the corresponding mode structure is shown in
Figs 1 a,b,c. As can be seen, the pump modes have decreased their amplitudes
by a factor of the order of $10^{6}$, from the interaction region to the end
of the filtering region (to the left), whereas mode 3 has essentially the
same amplitude in both regions. By increasing the filtering distance, the
pump signals could of course be reduced further. Naturally a filtering
geometry will make the mode coupling somewhat weaker, as compared to our
pure cylinder example. By adding a filtering region of roughly the same size
as the coupling region, the coupling factor reduces to $K_{\mathrm{fil}}\sim 
$ $K_{\mathrm{cyl}}/\sqrt{2}$, in which case we keep the same number of
excited photons if we choose the size of the total cavity region to be
roughly twice the size of the cylinder cavity presented in the previous
section.

\section{Nonlinearities in the walls of the cavity}

To our knowledge, a well-established and simple nonlinear model for the
superconducting RF-state does not exist. As a starting point for a
discussion, we may adopt a model with a nonlinear magnetic third order
susceptibility $M^{i}=\chi ^{ijlkl}B_{j}B_{k}B_{l}$, using the Einstein
summation convention, and defining the susceptibility in terms of $B$ rather
than $H.$ We then consider the same geometry of the cavity and the
eigenmodes as in section IV. The magnetic $z$-components of the pump fields penetrate roughly a
skin depth inside the walls. For a nonzero value of $\chi ^{3333}\equiv \chi
_{\mathrm{nl}}$, the part of the nonlinear magnetization that can act as a
source for mode 3 is then $\mathbf{M}_{3}=\chi _{\mathrm{nl}%
}B_{1z}^{2}B_{2z}^{\ast }\widehat{\mathbf{z}}$. Acoordingly we get currents
in the azimuthal direction $J_{3\varphi }\widehat{\mathbf{\varphi }}=\nabla
\times \mathbf{M}_{3}$, which in turn may induce radial variations in the
magnetic field $B_{3z}(\rho )=(\mu _{0}/\rho )\int J_{3\varphi }\rho d\rho $%
. At the same time, the jump in the nonlinear susceptibility across the
vacuum/superconductor boundary causes a jump in the magnetization, and
thereby a surface currrent $\ \mathbf{j}_{3s}=-\mathbf{M}_{3}\times \widehat{%
\mathbf{\rho }}$. A combination of the surface and bulk currents then yields a
nonlinearly induced magnetic field inside the vacuum region 
\begin{equation}
\left. B_{3z}\right| _{\rho <a}=(\mu _{0}/\rho )\int_{\lambda
}^{a}J_{3\varphi }\rho d\rho -\left. \mu _{0}\chi _{\mathrm{nl}%
}B_{1z}^{2}B_{2z}^{\ast }\right| _{\rho =a}  \label{Bondary-value}
\end{equation}
where $\lambda $ should be much larger than the skin depth such that $%
J_{3\varphi }(\lambda )$ is negligible. Carrying out the integration in Eq. (%
\ref{Bondary-value}), we see that the two terms cancel such that $\left.
B_{3z}\right| _{\rho <a}=0$, i.e. the nonlinear currents do not induce a
magnetic field inside the cavity. \

We stress that this is not a proof that nonlinearities in the walls
generally are unable to affect the physics inside the cavity. However, it
suggests that a nonlinear current in the walls with the proper
eigenfrequency does not by necessity excite the corresponding eigenmode of
the cavity.

\section{Discussion and Summary}

In the present paper we have materialized our proposal for the detection of
elastic photon-photon scattering \cite{Brodin-Marklund-Stenflo}. In
particular, we have calculated the output level of scattered photons in
terms of the allowed pump field strength and the cavity quality factors for
a reactangular prism as well as for a cylindrical geometry. Furthermore, we
have made finite element calculations to show that the resonance and
frequency matching conditions can be fulfilled in a filtering geometry,
where only the scattered mode has a high enough frequency to reach the
cavity region with a lower cross-section. Using performance data from
current state-of-the-art superconducting niobium cavities, where a high
quality factor $Q\approx 4\times 10^{10}$ is combined with surface field
strengths close to the critical value $B\approx 0.28$ $\mathrm{Tesla}$, we
deduce that the number of scattered photons can reach $N\approx 18$ in a
cylindrical cavity with length 2.5 m and 25 cm radius. In a filtering
geometry we estimate that the number of photons/volume will be reduced by a
factor of order 2, as compared to the pure cylinder case. Recent results 
\cite{Q-factor1,NOGUES99}, where measurements involving transitions in
Rydberg atoms interacting with single microwave photons have been made,
strongly suggest that the estimated field levels are within the range
detectable with present day technology.

We note that, in principle, nonlinearities in the walls of the cavity may lead
to excitation of the same mode as caused by the QED nonlinearities. Our simple model
calculation in section VI suggests that the mode-coupling due to such an effect
vanishes. However, a more rigorous treatment is necessary to draw definite
conclusions. QED theory predicts a definite output level of mode 3. 
Experiments that result in a much higher level would thus indicate 
that nonlinearities in the walls play the main role.

Finite element analysis, may provide a starting point for a suitable design
of the cavity. However, the degree of fine tuning of the resonance
frequencies of the modes is very high, since for optimal performance the
mismatch of the eigenfrequencies should not exceed $\delta \omega \sim $ $%
\omega /Q$. Hence the final adjustments of the cavity geometry must be made
experimentally.


\newpage

\begin{widetext}
\

\begin{figure}[ht]
\subfigure[]{\includegraphics[height=5cm,width=.85\textwidth]{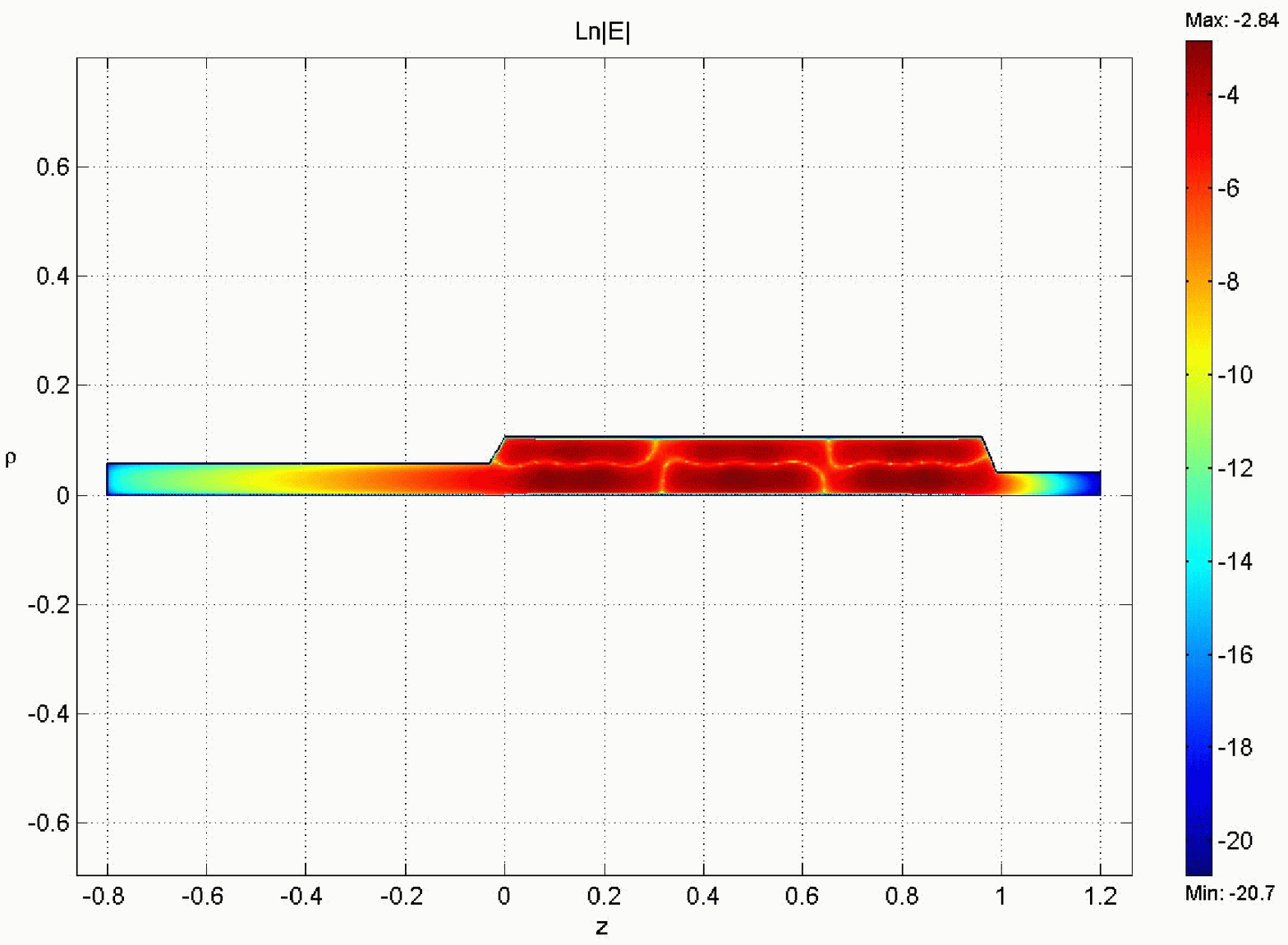}}
\subfigure[]{\includegraphics[height=5cm,width=.85\textwidth]{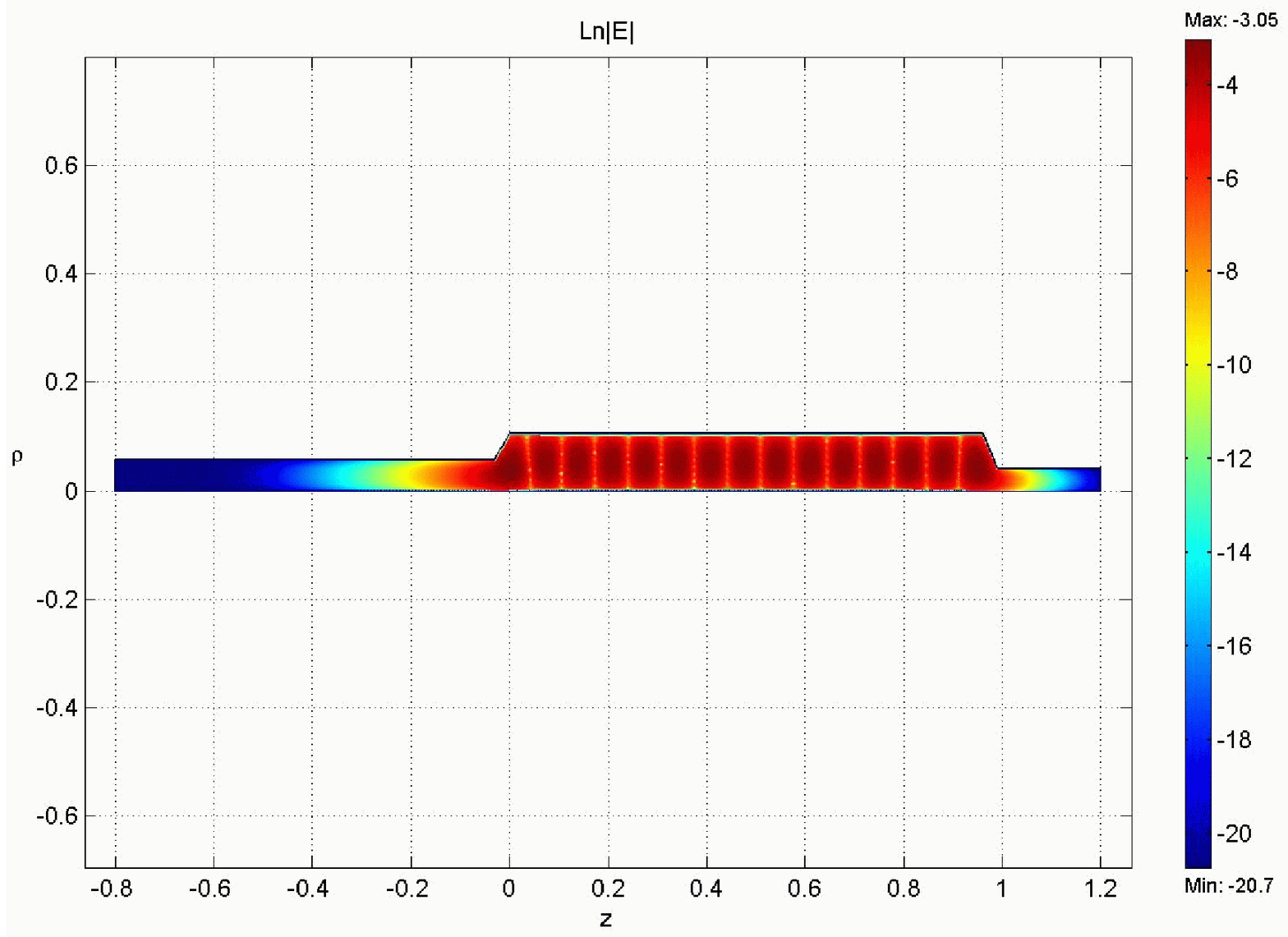}}
\subfigure[]{\includegraphics[height=5cm,width=.85\textwidth]{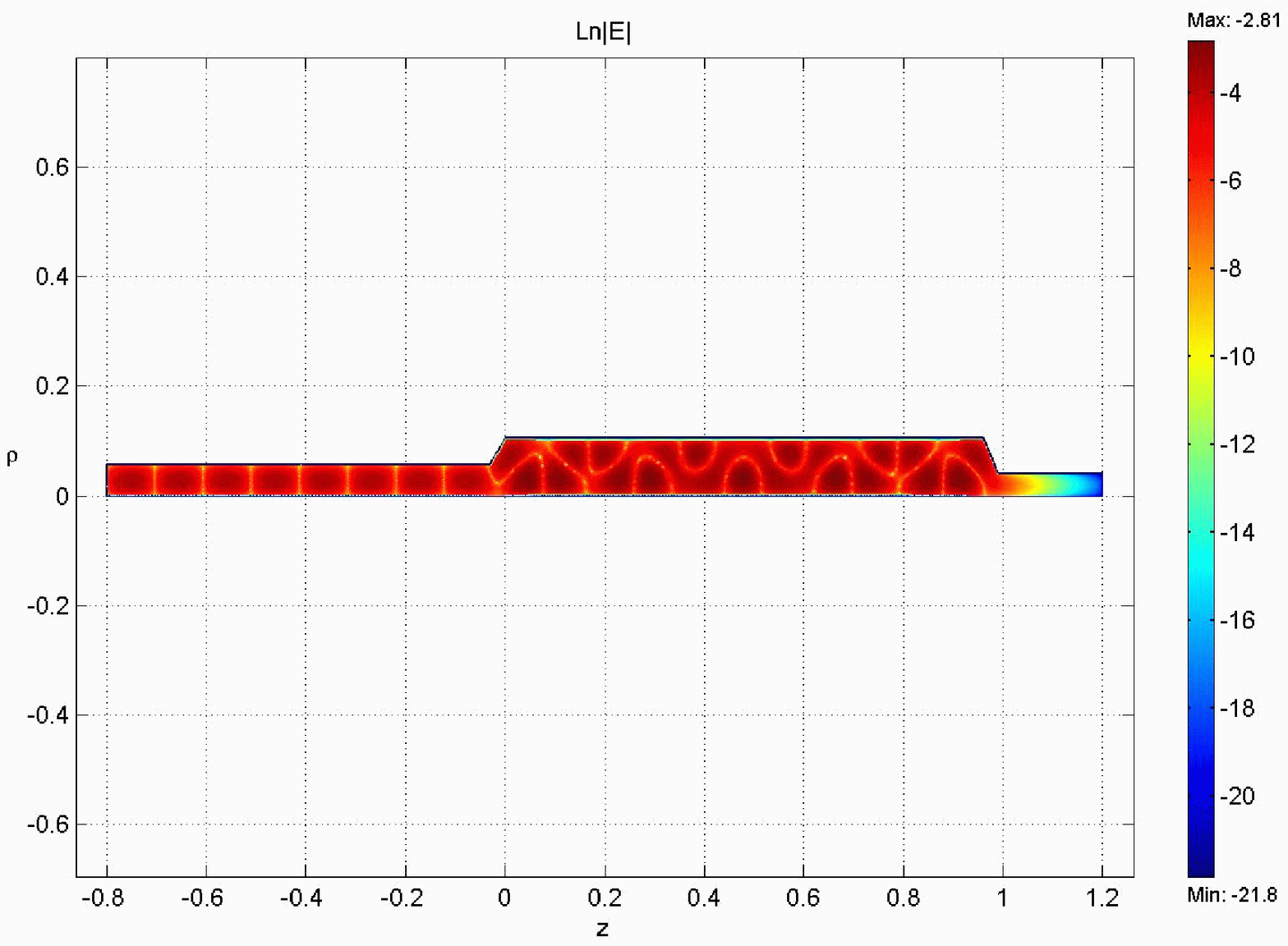}}
\caption{The geometry of the cavity and the mode structure for the filtering
geometry. Note that only half of the cavity is shown, since the other half
is redundant due to rotational symmetry of the cavity as well as the fields.
The small region to the right is the entrance region, the large middle
region is where the interaction takes place, and the region to the left is
the filtering region. All modes have the electric field in the angular
direction, i.e. $\mathbf{E}=E(\rho ,z)\widehat{\mathbf{\varphi }}$. The
variations of $\ln \left| E(\rho ,z)\right|$ are shown in color
code. \\
a) The mode structure of pump mode 1. The exponential decay in the
region of small cross-section diminishes the amplitude by a factor 
$10^{-6}$ in the end of the filtered region. \\ 
b) The mode structure of pump mode 2. The exponential decay in
the region of small cross-section diminishes the amplitude by a factor  
$10^{-8}$ in the end of the filtered region. \\
c) The mode structure of the excited mode. The amplitude is
roughly the same in the region of interaction and in the filtered
region.}
\end{figure}

\
\end{widetext}

\end{document}